\title {How Quantum Computers Can Fail}
\author {  Gil Kalai\footnote{Research supported in part by an NSF grant, 
by an ISF Bikura grant, and by a BSF grant. 
I am grateful 
to Dorit Aharonov, Michael Ben-Or, Greg Kuperberg, and John Preskill for 
fruitful discussions 
and to many colleagues for helpful comments.
}
 \\
Hebrew University of Jerusalem and Yale University}
\date {}
\documentstyle[12pt,amsfonts]{article}
\newcommand{\beq}[1]{\begin{equation}\label{#1}}
\newcommand{\enq}[0]{\end{equation}}

\renewcommand{\baselinestretch}{1.21}

\begin{document}

\maketitle

\begin {abstract}
We propose and discuss two postulates on the nature of errors in
highly correlated noisy physical stochastic systems. The first postulate 
asserts that errors for a pair of  
substantially correlated elements
are themselves substantially correlated. The second postulate asserts that 
in a noisy system with many highly correlated elements there will 
be a strong effect of error synchronization. These postulates 
appear to be damaging for quantum computers.

\bigskip

\bigskip

\bigskip

\end {abstract}

\newpage

\section {Quantum computers and the threshold theorem}

Quantum computers are hypothetical devices
based on quantum physics. A formal definition of
quantum computers 
was pioneered by
Deutsch \cite {D}, who also realized that they can outperform
classical computation.  The idea of a
quantum computer can be traced back to works by
Feynman, Manin, and others, and this development is also related to
reversible computation and connections between computation and physics 
that were studied by Bennett in the 1970s.
Perhaps the most important result in this field and
certainly a major turning point was
Shor's discovery \cite {S1} of a polynomial quantum
algorithm for factorization. The notion of a quantum computer 
along with the associated complexity class BQP
is an exciting gift from physics to
mathematics and theoretical computer science, and has generated
a large body of research. Quantum computation is also a source 
of new, deep, and unifying 
questions in various areas of experimental and theoretical physics.
For background on quantum computing, see
Nielsen and Chuang's 
book \cite {NC}.

Of course, a major question is whether quantum computers are
feasible. An early critique of quantum computation (put forward in
the mid-90s by Unruh, Landauer, and others) concerned the matter
of noise:

\begin {itemize}
\item [{\bf [P0]}]
{\bf The postulate of noise: Quantum systems are noisy.}
\end {itemize}

The foundations of noisy quantum computational 
complexity were laid by Bernstein and Vazirani in \cite {BV}.
A major step in showing that noise can be handled was the
discovery by Shor \cite {S2} and Steane \cite {St} 
of quantum error-correcting codes.
The hypothesis of fault-tolerant quantum computation (FTQC) was
supported in the mid-90s by the ``threshold theorem'' \cite {AB2,K1,KLZ,Got}, 
which asserts that under certain natural assumptions of statistical 
independence on the noise, if the rate of noise (the amount of noise
per step of the computer) is not too large, then FTQC is possible. It
was also proved that high-rate noise is an obstruction to FTQC.
Several other crucial requirements for fault tolerance were also described in
\cite {AB1,ABIN}.

The study of quantum error-correction and its limitations, as well as 
of various approaches to fault-tolerant quantum
computation, is extensive and beautiful; see, e.g., \cite
{CS,K2,K3,Kn,ND,Ra,BCLLSU}.  
 Concerns about noise models with statistical dependence are 
mentioned in several places, e.g.,
\cite {Pr,Le}. 
Specific models of noise that may be 
problematic for quantum error-correction are studied in \cite
{AHHH}.
Current FTQC methods apply even to more general models of noise
than those first considered, which allow various forms of time-
and space-statistical dependence; see \cite {TB,AGP,AKP}.

The basic conjecture of this paper is that
noisy highly correlated data cannot be stored or manipulated. 
On a heuristic level this conjecture is interesting for both the quantum 
and the classical cases.\footnote {Note that in
the classical case correlations do not increase the computational
power. When we run a randomized computer program, the random bits can 
be sampled once they are created, and it is of no computational 
advantage in the classical case to ``physically maintain'' highly correlated 
data.} The formal conjectures are restricted to the quantum case and refer to 
decoherence, namely the information loss of quantum systems.

Section \ref {s:n} gives more background on noise and fault-tolerance.  
An informal description of our conjectures in Section \ref {s:p}
is followed by a mathematical formulation in Section \ref {s:mf}.
Section \ref {s:d} is devoted to a discussion of  
the consistency of our conjectures with quantum mechanics and with the reality
of classical error-correction and fault tolerance. We also discuss 
the challenge of finding concrete noise models satisfying our conjectures, 
as well as connections with computational complexity theory and with physics.

\section {Noise and fault tolerance}
\label {s:n}


The postulate of noise is essentially a hypothesis about
approximations. The state of a quantum computer can be
prescribed only up to a certain error. For FTQC there
is an important additional assumption on the noise, namely on the
nature of this approximation.  The assumption is that the noise is
``local.'' This condition asserts that the way in which the 
state of the computer changes between 
computer steps is statistically independent, for different qubits. 
We will refer to such changes as ``qubit errors.'' 
In addition, the gates that carry the computation itself are imperfect.
We can suppose that every such gate involves at most 
two qubits and that the gate's
imperfection 
can take an arbitrary form, 
so that the errors (referred to as ``gate errors'') created on the two qubits 
involved in a gate can be statistically dependent. (Of course, qubit 
errors and gate errors propagate along the computation and handling this
is a main difficulty in fault tolerance.) 

The basic picture we have of a noisy computer is that 
at any time during the computation 
we can approximate 
the state of each qubit only up to some small error term 
$\epsilon$. Nevertheless, under the assumptions concerning the errors 
mentioned above, computation is possible. The noisy physical qubits
allow the introduction of logical ``protected'' qubits which are 
essentially noiseless.

The close analogy between the classical case and the quantum case for error
correction and fault tolerance is very useful.
For our purposes, a good way to understand 
the notions of quantum error-correction
and fault tolerance is to draw the line not between classical and
quantum information but between  deterministic information (or
even stochastic information where the elements are statistically
independent) and stochastic highly correlated information (both
classical and quantum). Thus, while the state of a digital
computer having $n$ bits is a string of length $n$ of zeros and ones,
in the (classical) stochastic version, the state is 
going to be a (classical) probability distribution
on all such strings. 

Quantum computers are similar to these (hypothetical) stochastic 
classical computers and they work on qubits (say $n$ of them). The state of 
a single qubit $q$ is described by a unit vector $u = a|0>+b|1>$  in 
a two-dimensional complex space $U_q$. 
(The symbols $|0>$ and $|1>$ can be thought of as 
representing two basis elements in $U_q$.) 
We can think of the qubit $q$ as representing 
$'0'$ with probability $|a|^2$ and $'1'$ with probability $|b|^2$. 
The state of the entire computer is a unit vector in the $2^n$-dimensional 
tensor product of these vector spaces $U_q$'s for the 
individual qubits. The state of the computer thus 
represents a probability distribution on the $2^n$ strings
of length $n$ of zeros and ones. The evolution of the quantum 
computer is via ``gates.'' Each gate $g$ 
operates  
on $k$ qubits, and we can assume $k \le 2$. 
Every such gate represents a 
unitary operator on $U_g$, 
the ($2^k$-dimensional) tensor product 
of the spaces that correspond to these $k$ qubits. 

A simple  (rather special) example of noise 
to keep in mind is that all qubit errors are 
independent random unitary operators for the individual qubits,
and all gate errors are 
random unitary operators on the spaces $U_g$. 
If these errors are 
small (namely, if all these operators are sufficiently close
to the identity), the threshold theorem will apply.

A main insight of quantum error-correction is that errors 
affecting a substantial but small fraction of --- even highly correlated --
bits/qubits can be handled. (For this, basic linearity properties 
of probability theory as well as of quantum physics are crucial.)
Errors that exceed, with
substantial probabilities, the capacity of the error-corrector are
problematic. Under the independence assumptions of 
the threshold theorems, if the 
rate of errors is small the probability for  
exceeding the capacity of the error-corrector is extremely small.
The crux of the matter is whether independent (or
almost independent) errors on highly correlated elements 
is a possible or even a physically meaningful notion.

\section {Noisy stochastic correlated physical systems}
\label {s:p}
\subsection {The postulate of noisy correlated pairs}
\label {s:p1}

The purpose of this section is to propose and discuss the
following postulate:

\begin {itemize}
\item [{\bf [P1]}]
In any noisy physical system, 
the errors for a pair of elements
that are substantially statistically dependent
are
themselves substantially
statistically dependent.
\end {itemize}
In particular, for quantum computers\footnote{Our conjectures themselves 
come in (highly correlated) pairs. Each conjecture is formulated 
first for general noisy physical systems and 
then specified to quantum computers, which are physical devices
able to maintain and manipulate highly entangled qubits.}
this postulate reads:

\begin {itemize}
\item [{\bf [P1]}]
In a quantum computer, 
the errors for a pair of substantially
correlated
qubits are substantially 
correlated.
\end {itemize}

Another way to put Postulate [P1] is: noisy correlated elements
cannot be approximated up to almost independent error terms: if
we cannot have an approximation better than a certain error rate
for each of two correlated elements, then an uncorrelated or almost
uncorrelated approximation is likewise impossible.


{\bf Remarks:}

1. {\bf Real-life examples: The weather 
and the stock market. }
We can discuss Postulate [P1] for cases of (classical) stochastic
systems with highly correlated elements.  I am not aware of a case of
a natural system with stochastic highly correlated elements that
admits an approximation up to an ``almost independent'' error term.
This is the kind of approximation required for fault-tolerant
quantum computation.

Can we expect to estimate the distribution of prices of two very
correlated stocks in the stock market up to an error distribution
that is almost independent?

Or take, for example, the weather. Suppose you wish to forecast
the probabilities for rain in twenty nearby locations. We
suppose these probabilities will be strongly dependent. Can we
expect to have a forecast that is off by a substantial error
that is almost statistically independent for the different locations?

To make this question a little more formal, consider not how accurately
a weather forecast  
predicts the weather,  
but rather how it  
predicts  (or differs from) a later weather forecast.
Let $\cal D$ be the distribution that represents the best forecast we
can give for the rain probabilities at time $T$ from the data we
have at time $T-1$. Let $\cal D'$ be the best forecast from
data we have at time $T-1-t$. 
Suppose that $\cal D$ is highly
correlated. Postulate [P1] asserts that we cannot 
expect that the difference ${\cal D} - {\cal
D'}$ will be almost statistically independent for two locations
where $\cal D$ itself is substantially correlated.

2. {\bf The threshold theorem and pair purification.} The
threshold theorem that allows FTQC has various remarkable
applications, but our postulate can be regarded as challenging its
simplest nontrivial consequence. The assumptions of 
the threshold theorem allow the errors on a pair of qubits 
involved in a gate to be statistically dependent. In
other words, the outcome of a gate acting on a pair of qubits 
prescribes the position of the
two qubits only up to an error that is allowed to exhibit an
arbitrary form of correlation. The  
process of fault tolerance
allows us to reach pairs of entangled qubits that, while still being
noisy, have errors that are almost 
independent.
This ``purifying'' nature of fault tolerance for quantum computation 
is arguably an element
we do not find in fault tolerance for deterministic
computation. 

3. {\bf Positive correlations for errors.}
Consider a noisy 
classical computer on $n$ bits 
and suppose that the overall error is given by taking the XOR of the $n$ 
bits in the computer with a randomly chosen string $e$ of $n$ bits 
according to 
a probability distribution $\cal E$. Suppose that for every bit, the error 
probability is 1/1000. If the errors are independent then the probability 
that $e$ will have, say, $n/500$ error is very tiny as $n$ grows. 
Positive correlations between the errors 
for all (or most) pairs of bits will change this picture.
For example, if for every two bits the 
probability that for $e$ both these bits are 1 is 
around 1/50,000 (rather than $10^{-6}$), then there will be a substantial 
probability that more than $n/100$ bits will be ``hit'' by the error.
(The same conclusion will apply if 
for every triple of bits
the probability that they will all be hit 
is, say, $10^{-7}$ rather than $10^{-9}$.) 
This effect of positive correlation for errors is the 
basis for Postulate [P2] below.

4. {\bf Leaks of information.}
Rather than talk about errors and noise we can talk about 
information ``leaked'' from our physical systems 
to the outside world. For quantum computers 
leaking of information automatically amounts to noise and thus a strong form 
of Postulate [P1] for quantum computers is:

\begin {itemize}
\item [{\bf [P1']}]
For a noisy quantum computer, 
information leaks for two substantially
correlated
qubits have a substantial positive 
correlation.
\end {itemize}

For general stochastic systems [P1'] reads:

\begin {itemize}
\item [{\bf [P1']}]
In any noisy physical system, the information leaks
concerning the states of two  elements
that are substantially statistically dependent
have a substantial positive correlation.
\end {itemize}

Postulate [P1'] seems natural
for systems where correlations are gradually created and 
information is gradually leaked. The central question  
is whether such an effect can be diminished 
via error correction.

\subsection {The postulate of error synchronization}
\label {s:es}

Suppose we have an error rate of $\epsilon$. The assumptions of
the various threshold theorems (and other proposed methods for 
quantum fault-tolerance) imply that the probability of a
proportion of $\delta$ qubits being ``hit'' is exponentially small
(in the number of bits/qubits)
when $\delta$ exceeds $\epsilon$. Error synchronization refers to an
opposite scenario: there will be a substantial probability of a
large fraction of qubits being hit. 


\begin {itemize}
\item [{\bf [P2]}]
In any noisy physical system with many substantially correlated
elements there will be a strong effect of spontaneous error-synchronization.
\end {itemize}

\begin {itemize}
\item [{\bf [P2]}]
In any quantum computer at a highly entangled state there will be a
strong effect of spontaneous error-synchronization.
\end {itemize}


As remarked above, error synchronization is expected for a large system 
when the errors (or information leaks) are   
positively correlated.
An even stronger form of error synchronization is 
considered in \cite {Ka}, where  formal 
definitions for the quantum case can be found.

{\bf Remarks:}

1. {\bf Empiric.} Postulates [P1] and [P2] can be tested, in
principle, for quantum computers with a small number of qubits
(10-20). Even if such devices where the qubits themselves are
sufficiently stable are still well down the road, they are  
to be expected long before the superior complexity power of quantum computers 
kicks in. 

The rigorous form of Postulate [P1] (Section \ref {s:mf}) 
can be suggested as a benchmark 
for quantum-computer engineers: to construct pairs of noisy
entangled qubits with almost independent error-terms.

2.  {\bf Spontaneous synchronization for highly correlated systems.} The
idea that for the evolution of highly correlated systems changes
tend to be synchronized, so that we may witness rapid
changes affecting large portions of the system (between long
periods of relative calm), is appealing and may be related to other 
matters like sharp
threshold phenomena and phase transition, the theory of evolution, 
the evolution of
scientific thought, and so on.\footnote{ This idea is conveyed
in the Hebrew proverb ``When troubles come they come together.''}
We can examine the possibility of error
synchronization for the examples considered above.
Can we expect synchronized errors for weather forecasts? Can we
expect stock prices, even in short time scales, to exhibit
substantial probabilities for changes affecting a large proportion
of stocks? 
Spontaneous synchronization is also related to the issue of pattern
formation for correlated systems.

3. {\bf Error synchronization and the concentration of measure
phenomenon.} A mathematical reason to find spontaneous
synchronization of errors an appealing possibility is that  it
is what a ``random'' random noise looks like. Talking about a random
form of noise is easier in the quantum context. If you prescribe
the noise rate and consider the noise as a random (say unitary)
operator (conditioning on the given noise rate), you will see a perfect
form of synchronization for the errors, and this property will be
violated with extremely low probability. 

Random unitary operators
with a given noise rate are {\it not} a realistic form of noise. 
The qubits in a quantum computer are expected to be quite isolated, 
so that the errors are described by a ``locally defined'' process 
(namely, a process (stochastically) 
generated by operations on a small number of qubits at a time) --- not unlike
the (noiseless) evolution described by quantum computation itself. 

While random unitary operators with a prescribed error rate 
appear to be unapproachable by any
process of a
 ``local'' nature, the issue is whether some of their statistical properties 
may hold for such stochastic
processes describing the errors. 
The fact that perfect error-synchronization 
is the ``generic'' form of noise may suggest that
stochastic processes describing the noise will approach this
``generic'' behavior unless they have  good reason
not to. (One obstruction to error synchronization, pointed out 
by Greg Kuperberg, is time independence.)

4. {\bf Correcting highly synchronized errors.} An observation
that complements the discussion so far is that synchronized
errors that are unbiased 
can be corrected to produce noiseless
deterministic bits. Suppose we have a situation in which an error hits
every bit with probability $(1-\epsilon)$ and when a bit is hit it
becomes a random unbiased bit. (That is, a
bit is flipped with probability $(1-\epsilon)/2$.) This type of
noise can be corrected by representing a 0 bit by a long string
of 0's and a 1 bit by a long string of 1's. (If the noise hits a smaller 
fraction of bits, the condition of it being unbiased 
can be compromised.) If we start with qubits and again, with probability $(1-\epsilon)$, replace each qubit 
with a random (uniformly distributed) qubit, 
we can still
extract noiseless {\it bits}.
However, there is no quantum error-correction code for such noise.

This means that deterministic noiseless
bits can prevail (for classical and quantum systems) 
even for some forms of highly correlated errors.
(Our postulates do not imply a high correlation for the errors 
when the elements of the system are statistically independent, but
mechanisms leading to our conjectural effects may still be
relevant for the nature of noise 
for  classical forms of
storing information and computation.)

The method of ``clone and sample'' 
appears to be essentially the only error-correction method we find in nature.
This method allows us to introduce gates where errors on the involved bits 
will be almost independent to start with, and thus will 
reduce ``noise on gates'' to ``noise on bits.'' But this method
is unavailable for quantum information of a general type. 

5. {\bf The censorship conjecture}. Notions of ``highly correlated''
or ``highly entangled'' systems are not easy to define. We will
refer informally to systems that up to a small error are induced
by their marginal distributions on small sets of elements as
``approximately local.''

\begin {itemize}
\item [{\bf [C]}]
Censorship conjecture: Noisy stochastic physical systems are approximately
local.
\end {itemize}

\begin {itemize}
\item [{\bf [C]}]
The states of quantum computers are approximately local.
\end {itemize}

The rationale for this conjecture is that high forms of
entanglement will necessarily be accompanied 
with 
a strong effect of error synchronization,
which in turn will push the system towards approximate 
locality. 

A suggested definition of
``approximately local'' (for the quantum case only), and a precise
formulation of the conjecture are given in the next section 
(see also \cite {Ka} for a different approach).\footnote{There are 
many measures for the ``amount of entanglement'' 
(and correlation) that can be used. It is 
also unclear if we should measure the entanglement of a single state or 
use a measure that depends on the variety of feasible states for a system. 
Leggett's early paper \cite {L} and 
his ``disconnectivity measure'' (D-measure) 
seem relevant.}

\section {A mathematical formulation} 
\label {s:mf}

\subsection {Measuring information leaks}

In this section we give a mathematical 
formulation for Postulate [P1'] and present even stronger conjectures. 
Our setting is as follows. We have a quantum computer running on $n$ qubits.
The noise can be described
by a unitary operator on the computer qubits and the neighborhood qubits
or as a quantum operation $E$ on the space of 
density matrices for these $n$ qubits. The ideal state of the 
quantum computer is pure.

Consider the conjectures in this section 
in the following way. A noisy quantum computer is subject 
to noise described by 
a quantum operation $E$, such that the error rates for individual qubits
are small but substantial and $E$ satisfies the 
requirements described in this section. 
The operation $E$ need not be 
the overall noise that describes the gap between the ideal state 
and the noisy state of the computer, but we assume that any 
damaging properties
of $E$ will not be remedied by additional noise of a 
different nature. 

We denote by $L(a)$ the ``amount of information the 
neighborhood has on the qubit $a$.'' More generally, for a set $A$ of 
qubits we denote by $L(A)$ the ``amount of information the 
neighborhood has on $A$.'' Next we propose mathematical definitions 
for these notions.

Let $\rho$ be a state of the computer. For a set $A$ of qubits
let $\rho|_A$ be the induced state on $A$.
When the state $\rho$ is a tensor product pure state  
then for every set $A$
of qubits, $S(\rho|_A)=0$ and the information leak of 
the noise operator $E$
from the set of qubits $A$ can be measured by 
the entropy $S((E \circ \rho)|_A)$. 
Here, $S(*)$ is the (von Neumann) entropy function 
see, e.g., \cite {NC}; Ch. 11.
(We deem this entropy-based 
notion appropriate for our purposes, even though the entropy does not capture
every form of ``information leak'' attributable to a noise 
operator.)

I am unaware of any canonical 
way to make the ``information leak'' a measure
of the noise operation $E$ that does not depend on a specific 
choice for this tensor product state. In what follows 
we let $\rho_0 = (+)^{\otimes n}$ $= (1/ \sqrt 2 ((|0> + |1>))^ {\otimes n}$ 
and define $L(A)=L_E(A)=S(E(\rho_0|_A))$.

{\bf Remark:} Let $\hat \rho$ be the state of the computer's qubit and the 
environment that is represented by a set $N$ of qubits. A standard measure
of the information that the environment has on the qubits in $A$  
is $L'(A) = S(\hat \rho|_A) + S(\hat \rho|_N) - S(\hat \rho|_{\{A\cup N\} })$.
I would expect that $L'(A)$ can replace $L(A)$ for the formulation 
of the conjectures in this section.

\subsection {Two qubits}

We will state mathematically a version of Postulate [P1'].
Our setting is as follows. Let $\rho $ be the ``ideal'' state 
of the computer and consider two qubits $a$ and $b$. 
We  use as the (rather standard) measure of entanglement

$$ ENT(\rho; a,b)= S(\rho |_a)+S(\rho |_b) - S(\rho |_{ \{a,b\} }).$$
\noindent
As a measure of correlation of information leaks we use
$$EL(a,b) = L(a)+L(b)-L(\{a,b\}).$$

Postulate [P1'] can be formulated as:

\begin {equation}
\label {e:p1}
EL(a,b) \ge  K(L(a),L(b)) \cdot ENT(\rho; a,b),
\end {equation}

\noindent
where $K$ is a function of $L(a)$ and $L(b)$,  
which is substantially larger than 
their average $(L(a)+L(b))/2$. ($K (0,0)=0$, so that 
relation (\ref {e:p1}) tells us nothing about 
noiseless entangled systems.)


{\bf Remark:} We are mainly interested in the case 
where the error rate is fixed, 
but the dependence of $K (L(a),L(b)) $ on the error rates 
is also of interest.


\subsection {Two qubits: A stronger version}

We go on to describe and motivate an even stronger form of 
[P1'] and an extension to more than two qubits. These extensions
go beyond Postulates [P1] and [P2] as discussed so far.

The expression $S(\rho|_a)+S(\rho|_b) - S(\rho|_{\{a,b\}})$ was used 
as a measure of entanglement between two qubits. We would like 
to replace it by a measure that can be 
called ``emergent entanglement,'' which we are now going to define.
This measure, denoted by $EE(\rho; a,b)$, captures (roughly)  
the expected amount of entanglement among the two qubits 
when we measure some other qubits, ``look at the outcome,'' and condition
on all possible outcomes for the measurement.
It appears to be related to Briegel and Raussendorf's 
notion of ``persistent entanglement'' \cite {BR}. 




For every representation $\omega$ of $\rho|_{ \{ a,b \} }$
as a mixture (convex combination) of 
joint states 
$$\rho|_{ \{ a,b \} } = \sum_{i=1}^t p_k \rho _k,$$ let 
$$ENT_\omega (\rho ;a,b ) = \sum p_k ENT(\rho_k;a,b).$$  
Define $$EE(\rho; a,b) = \max ENT_\omega (\rho; a,b), $$ where the maximum
is taken over all representations $\omega$. (We can assume that $\omega$ 
is a mixture of pure joint states.)

A strong form of relation (\ref {e:p1}) is
\begin {equation}
\label {e:p1strong}
EL(a,b) \ge  K(L(a),L(b)) \cdot EE(\rho; a,b),
\end {equation}

\noindent
where, as before, $K$ is a function of $L(a)$ and $L(b)$  
which is substantially larger than 
their average $(L(a)+L(b))/2$. 





The motivation for this strong version of Postulate [P1'] comes from 
considering the state of a quantum computer that applies 
a fault-tolerant computation.
The state of the computer is $t$-wise independent for a large value 
of $t$; hence every two qubits are statistically independent and 
Postulate [P1'] does not directly apply.
Consider an error-correcting code and let $s$ be the minimal number 
of qubits whose state ``determines'' that of the others, so that once 
they are measured and their value are ``looked at'' the state of the 
other qubits is determined. When we measure and look at the values 
of $s-1$ qubits, we see a very strong dependence between 
every pair of the remaining qubits. Now, if we assume Postulate [P1']
and (just tentatively) also assume that ``measuring and looking at'' 
the contents of some qubits 
does not induce errors on other qubits (this is a standard assumption 
in current noise models), we see that the conclusion 
of Postulate [P1'] should apply for pairs of qubits in a 
quantum computer running FTQC even though pairs of qubits are independent.


\subsection {More qubits}

Here is a suggestion for an extension of the above 
conjecture from pairs of qubits to  larger sets of qubits. 
This suggestion goes beyond Postulates [P1] and [P2] 
and is related to a strong form of 
error synchronization conjectured in \cite {Ka}.


For a set $A = \{a_1,a_2,\dots,a_m\} $ of $m$ qubits let 

$$ENT(\rho; A) = -S(\rho)+ \max S(\rho^*),$$ 
where $\rho^*$ is a mixed state with the same marginals
on proper sets of qubits as $\rho$, i.e., 
$\rho^*|_B = \rho|_B$ for every proper subset $B$ of $A$.

Define in a similar way $$EL(A) = -L_E(A)+L_{E^*}(A),$$
where $E^*$ is a quantum operation which satisfies
$E^*|_B=E|_B$ for every proper set $B$ of $A$.

Using these definitions we will extend our conjectures, given by 
relations (\ref {e:p1}) and (\ref {e:p1strong}), from pairs of qubits 
to larger sets of qubits. 
Let $\rho$ be an ideal state of the computer and 
let $A$ be a set of $m$ qubits. Extending (\ref{e:p1}) we conjecture that

\begin {equation}
\label {e:p1s}
EL(A) 
\ge  
K_m 
ENT(\rho|_A). 
\end {equation}

\noindent
Here, $K_m = K_m(\{L(a): a \in A\})$ is substantially 
larger than $\min \{L(a))):a \in A\}$ and it vanishes when all the 
individual information leaks vanish. 

Here again we  
further conjecture that for every representation $\omega$ 
of the state $\rho|_A$ 
as a convex combination $\rho|_A = \sum p_k \rho_k$ of pure
joint states,   

\begin {equation}
\label {e:p1sstrong}
EL(A) 
\ge  
K_m 
\sum p_k ENT(\rho_k; A). 
\end {equation}



{\bf Remarks:} 
1. We expect that a quantum error-correcting code that 
corrects $t$-errors and has a fixed error rate will have a strong form of 
error-synchronization as $t$ tends to infinity. Namely, the noise operation 
will have a similar effect to that of the following operation: 
with probability $\epsilon$ a $(1-o(1)$-fraction of qubits are 
being measured. (This is referred to as ``devastating'' noise
in \cite {Ka}.)  
I expect 
that Postulate [P1'] as expressed by 
relation (\ref {e:p1strong}) will imply 
the weaker form of 
error-synchronization  discussed in Section \ref {s:es}, while an 
extension for larger sets of qubits given by (\ref {e:p1sstrong})
will imply the stronger form.

2. The value of $ENT(\rho; A)$ is intended to 
serve as a measure of the additional information when we pass 
from ``marginal distributions'' on proper subsets of 
qubits to the entire distribution on all qubits. 

\subsection {Censorship}

Here is a suggestion for an entropy-based 
mathematical formulation for Conjecture [C]. We remind the readers that 
in this section  we always assume that the ``ideal'' state of the quantum 
computer (before the noise is applied)
is a pure state. Some adjustments to our conjectures 
will be required when the 
ideal state itself is a mixed state. 

Let $\rho$ be a pure state on a set 
$A = \{ a_1,a_2,\dots,a_n\} $ of $n$ qubits. Define 

$$\widetilde{ENT}(\rho) = \sum \{ENT (\rho; B): B \subset A \}.$$

\noindent
In this language a way to formulate the censorship conjecture is:

\medskip

\noindent
There is a 
polynomial $P$ (perhaps even a quadratic polynomial) such that 
for any quantum computer on $n$ qubits, 
which describes 
a pure state $\rho$, 
\begin {equation}
\label {e:c}
\widetilde{ENT} (\rho) \le P(n). 
\end {equation}

\medskip


We will mention now some mathematical challenges.
It will be interesting to prove relation (\ref {e:c}) based on
relation (\ref {e:p1s}), and to formulate and prove 
weak and strong forms of 
error synchronization based, respectively, on 
relations (\ref {e:p1}) and (\ref {e:p1s}).

A further goal would be to derive, based on the assumptions 
on noise for the physical qubits 
(relations (\ref{e:p1s}) and  (\ref{e:p1sstrong})), the same relations
as well as relation (\ref {e:c}),  
for  ``protected'' qubits, namely
logical qubits represented by quantum error-correction.

{\bf Remarks:} 
1. It is interesting to study 
how the quantities $ENT(A; \rho)$ 
evolve in time for 
dynamical systems describing (quantum and classical) physical processes.

2. The additional conjectures of this section 
are meant to draw the following picture: 
we have an ideal notion of a quantum computer that has 
extraordinary physical and computational properties.
Next come noisy quantum computers with an ideal notion of noise.
If the noise rate is small then FTQC is possible.
Next come noisy quantum computers that satisfy 
relation (\ref {e:p1}). For those, 
fault tolerance will require controlling the error rate as well as
$K_2$, which we expect to be much harder. This model
is also an idealization as long as $K_3=0$ and so on. For such 
highly entangled
states as those required in quantum algorithms, $K_i$ will be 
more and more damaging for larger values of $i$.





\section {Discussion}

\label {s:d}



Our  conjectures on the nature of noise 
for correlated systems
appear to be damaging to the possibility of storing and
manipulating correlated quantum or classical stochastic data and therefore
for the possibility of computationally-superior quantum computers.

In Section \ref {s:con} we ask if our conjectures are consistent with 
quantum mechanics. We also examine if they are  
consistent with the reality of 
classical error-correction and fault-tolerant classical 
computation. 
In Section \ref {s:mod} we ask if our  
conjectures can be supported by  concrete  
models of noise. In Section \ref {s:cs} we discuss the computational 
complexity consequences of the conjectures, and in Section \ref {s:phy}
we ask if there are any existing or expected 
counterexamples from physics.



\subsection {Consistency}
\label {s:con}

\subsubsection *{Causality}

We do not propose that the entanglement of the pair of noisy qubits 
{\it causes} the dependence between their errors. The correlation between
errors can be caused by the process leading to the correlation
between the qubits, or simply by the ability of the device to achieve
strong forms of correlation.  

\subsubsection *{Linearity}   Do our postulates violate the
linearity of quantum physics? The plain simple answer is no. Again
the analogy with classical stochastic processes is telling. The
conjecture that in noisy systems like the weather substantially
correlated events are subject to substantially correlated noise
(or, in other words, can only be approximated up to error terms
that are also substantially correlated) is perhaps bold and possibly
false, but it is not remotely bold enough to violate the laws of
probability theory. This is also so in the quantum case.

\subsubsection *{Probability, secrets, and computing}

We will now describe a 
difficulty for our conjectures at
least in the classical case. Consider a situation where Alice
wants to describe to Bob a complicated correlated distribution $\cal
D$ on $n$ bits that can be described by a polynomial-size
randomized circuit. Having a noiseless (classical) computation
with perfect independent coins, Alice can create a situation 
where for Bob the distribution of the $n$ bits is 
described precisely by $\cal D$. 
In this case the values of
the $n$ bits will be deterministic and $\cal D$ reflects Bob's
uncertainty. 
Alice can also make sure that for Bob the distribution of the $n$
bits  will be ${\cal D} + {\cal E}$, where $\cal E$ describes independent
errors of a prescribed rate.


Is this a counterexample to our Postulates [P1] and [P2]? One can
argue that the actual state of the $n$ bits is deterministic and
the distribution represents Bob's uncertainty rather than  ``genuine''
stochastic behavior of a physical device.\footnote {Compare the
interesting debate between Goldreich and Aaronson \cite {GA} on
whether nature can ``really'' manipulate exponentially long
vectors.} But the meaning of ``genuine stochastic behavior of a
physical device'' is vague and perhaps ill-posed. 
Indeed, what is the
difference between Alice's secrets and nature's secrets? In any case, 
the difficulty described in this paragraph cannot be 
easily dismissed.\footnote {The distinction
between the two basic interpretations of probability as either expressing
human uncertainty or as expressing some genuine physical
phenomenon is an important issue in the foundation of (classical)
probability. See, e.g., Anscombe and Aumann \cite {AA}. Opinions range from
not seeing any distinction at all between these concepts to
regarding human uncertainty as the only genuine interpretation.} 

However, note that as in the case of faraway qubits, the noisy 
distribution  ${\cal D} + {\cal E}$ was based on the ability to achieve
the noiseless distribution $\cal D$. Achieving the distribution $\cal D$ 
was based on noiseless classical computation to start with. 
For the case of quantum computers, we can still defend our 
Postulates [P1] and [P2] against 
this argument as follows. 
Even if nature can simulate Alice, and Bob's ``mental'' uncertainty 
can be replaced by a ``real'' physical 
situation where a 
highly
correlated distribution is prescribed up to an independent error
term, 
this approximation has been achieved via a noiseless
computation to start with. 
Therefore, such an approximation 
cannot serve, in
the quantum case, as a basis for fault tolerance.

The difficulties 
considered here motivated the formulation of our conjectures 
(Section \ref {s:p1}, remark 4, and Section \ref {s:mf}) 
in terms of information leaks.\footnote{A related idea, relevant 
also to the next item of faraway qubits, is  
%
to regard a stricter 
definition of a noisy quantum computer 
as such that 
at {\it any time} along the computation for every qubit,  
and {\it for every observer}
(who extracts information from the computer) 
the noise rate for every qubit (namely, the difference between its 
ideal state and its actual state) 
is at least $\epsilon$.} 
The mathematical formulation of our postulates in Section \ref {s:mf}
is thus restricted to the quantum case.\footnote{For the classical 
case (or for a commutative fragment of noncommutative probability) 
our postulates [P1] and [P2], as well as the postulate of noise [P0] and
even the notion of noise itself, 
are meaningful only on a heuristic or subjective level.} 


\subsubsection *{Faraway qubits }

Suppose we have two qubits that are far away from each other 
at a given entangled state at time $T$.
Consider their state at time $T+t$. Is there any reason to believe 
that the changes will not be independent? And if $t$ is small compared to the
distance between the qubits isn't it the case that to implement a noise 
that is not independent we will need to violate the speed of light? 
And finally isn't this observation a counterexample to Postulate [P1]? 

The answer to the last question is negative. There is no 
difficulty in conceding that changes over time 
in the states of two faraway entangled 
qubits will be independent. The problem with this critique 
is the initial assumption: we are {\it given} two qubits 
at time $T$ at a given state. Starting with noiseless 
correlated elements, we may well reach correlated elements 
that can be described up to substantial but independent error terms. 
But for fault tolerance we may not assume noiseless pairs 
of entangled qubits to start with. 

In this paper, as in other papers dealing with FTQC, 
we assume that at any time during the computation every qubit is noisy. 
Sometimes, a quantum computer that is only partially noisy is studied
(e.g., \cite {BCLLSU}). In such a case 
we should reformulate Postulates [P1] and [P2]
relative to the noiseless part.

\subsubsection *{Running a quantum algorithm with a ``random'' state 
at all times}

A critique of the possibility of any systematic 
damaging relation between the state 
of the quantum computer and the noise was suggested 
by Ben-Or (see \cite {Ka}) and is related 
to some works of Preskill and Shor. (A related concern was pointed out 
by Preskill. A detailed proof of such a result along 
with an interesting interpretation was recently offered by 
Aharonov \cite {Dorit}.) 
Having a classical computer control a quantum computer 
makes it  possible  to run a variant of any quantum computer program where at 
the initial state we apply  random Pauli operators on every qubit 
and modify the action of the gates accordingly. 
This 
interesting critique does not apply to the mathematical 
formulation given in Section \ref {s:mf} for Postulate [P1'], 
since
the  measures of entanglement we use
are invariant under such an operation.


To sum up this part of the discussion, 
our conjectures, properly formulated in terms of decoherence 
of quantum systems, are consistent
with the reality of classical fault-tolerance. They also 
appear to be fully consistent with quantum mechanics. On the other hand, 
as proposed properties of decoherence our conjectures 
are not in agreement with the common point of view regarding decoherence 
and the traditional decoherence models.


\subsection {Models}
\label {s:mod}





A  basic remaining challenge is to present concrete models of noise
that support Postulates [P1] and [P2]. (Of course, there is a 
difference
between showing that the type of behavior 
we are looking for is possible and showing that it is unavoidable.) 
A model for the noise that supports our postulates should already exhibit 
[P1] and [P2] for the ``new errors'' --- either qubit-errors 
or gate-errors or both ---
and would thus be quite different from current models. 

It is worth noting that error synchronization is a very familiar phenomenon
for error propagation of (unprotected) quantum programs.
It is instructive to see in this context how error synchronization is 
often created when we start with small independent 
errors and let them propagate along the steps of a computer program.



One way to view the noise is as represented by a rather
primitive (but quick) stochastic program (or circuit) ``running'' 
along the actual
program. We run the program $\cal P$ and we actually get ${\cal P}
+ {\cal N}$. The simplest explanation for why errors of 
correlated qubits are themselves correlated is that the noise
$\cal N$ depends on $\cal P$, or can be described as a weak
perturbation of the original program itself. But this is not the
only possibility. It may be the case that $\cal N$ does not 
depend on $\cal P$ but rather represents a certain 
form of ``generic'' quantum program. In both these cases we think of 
$\cal N$ as a quantum program with many steps for each computer cycle. 
This hypothetical ``noise program'' partially achieves one familiar 
``computational task'' for a distributed system: synchronization.


Recently, Klesse and Frank \cite {KF} described a physical system in which 
qubits (spins) are coupled to a bath of massless bosons. They 
reached (after certain simplifications) a noise model with 
error synchronization. (I am thankful to 
Robert Raussendorf for this information.) 
The earlier models suggested by Alicki, Horodecki, Horodecki, and Horodecki
\cite {AHHH} appear to be relevant to our conjectures. 
Also relevant is Alicki's
idea \cite {Al} (see also \cite {ALZ}) that ``slow gates'' (combined with 
the free evolution of the system) will be an obstacle to error correction.

There is a substantial interest
in local stochastic behavior leading to spontaneous (collective)
synchronization (e.g., \cite {KaSa,StS,DCMH,NRVB,Ku}). 
The Glauber dynamics (a very simple locally defined ``program'') 
for the Potts model (e.g., \cite {Mar}) 
can also be regarded in this way.
There is also a substantial amount of work on emergence of patterns 
in stochastic (correlated) locally described systems.

Finally, let me mention the relevance of {\it cluster states} defined by 
Briegel and 
Raussendorf (see, \cite {RBB}).  
(We will further discuss cluster states below.) 
The description of cluster states involves 
an array of qubits located on the vertices of a rectangular
lattice in the plane (or in space).
Cluster states are ``generated'' by 
local entanglements between pairs of nearby 
qubits on the lattice grid. 
They can be regarded as the quantum analogs of the Ising 
and Potts classical models. 
(Note that ``cluster state'' is a collective name for a large number
of possible models.) There is some evidence 
in the literature (see \cite {GRAC}) that cluster states 
emerge in realistic situations. 

Controlled creation and manipulation of cluster states can be
very important for building quantum computers. 
On the other hand, 
cluster states
(and decohered cluster states)
can serve as a basis
for concrete models of noise. 
Local processes leading (in reality) to 
cluster states may represent realistic 
models of decoherence. 
This possibility deserves further study. (I am thankful to 
Scott Aaronson for fruitful discussions concerning 
cluster states.) 


\subsection {Computation complexity}
\label {s:cs}

While it looks intuitively correct that our postulates are
damaging for quantum computation, proving it, and especially
proving a reduction all the way to the classical model of
computation, is not going to be an easy task. 
This is an interesting
question in computational complexity. In an earlier paper \cite {Ka}
some problems on the computational power of quantum computers with various
hypothetical types of  noise were considered. Reduction of low-rate noisy 
quantum computation to BPP is not known even for cases where the noise is 
provided by an adversary. A more realistic task would be to show 
that our postulates exclude 
fault tolerance based on linear quantum error-correction.

Let me first mention a few relevant earlier works.
The problem of describing complexity classes of quantum computers
subject to various  models of noise was proposed by Peter Shor
\cite {S3} in the 90s, but apparently was not picked up. See also 
\cite {Aa2}. It is even 
theoretically possible 
that deviating from the standard 
assumptions on noise (and, in particular, allowing 
dependence of the noise $\cal N$ on the program
$\cal P$),
will 
allow stronger computation power than
BQP. 

Aaronson \cite {Aa1} proposed a theory 
to study the computational-complexity effect of an arbitrary form
of restriction on states of quantum computers, namely, some states are 
forbidden while others can be freely prepared and manipulated.
Aaronson was motivated by  
several skeptical opinions on quantum computers, 
especially Levin's paper \cite {L}, and he proposed his theory 
as a way ``to make the debate concerning quantum computers more 
scientific and less ideologic.''  The approach taken here 
is close, in broad strokes, to Aaronson's. (A notable difference in the 
rhetoric is that Aaronson equates the failure of 
quantum computers with the breakdown of quantum mechanics.)


Going back to the issue of how damaging our postulates on noisy quantum 
computers can be, we note  
that going below the 
computation power of logarithmic depth polynomial-size 
quantum circuits appears to be difficult, 
yet such circuits combined with classical computers are
strong enough to allow a polynomial-time algorithm for 
factoring as follows from 
a recent result of Cleve and Watrous \cite {CW}.

Aharonov, Ben-Or, Impallazio, and Nisan \cite {ABIN} 
proved that the computational power of noisy {\it reversible} 
quantum computers reduces to log-depth quantum computation. 
But it is not even known  whether or not noisy reversible 
quantum computers (combined with  
a noiseless classical one), under the standard noise models, 
allow polynomial-time factoring, and it may well be the 
case that they do. Razborov \cite{Ra} showed that, for a certain (standard) 
model of 
noise, if the qubit-error rate is 50\% then the computation power
reduces to log depth quantum computation. In this case too, a reduction 
to classical computation is unknown. 
These results suggest that also in our case it will be difficult 
to prove a reduction 
which excludes  log-depth quantum circuits. Moreover, it is 
quite possible that log-depth quantum circuits prevail, with quasi-polynomial 
or even polynomial overhead, under 
rather general forms of noise. This is the case when the noise is described
by a random unitary operator; the perfect error-synchronization allows
log-depth circuits to work perfectly with a probability which is 
only polynomially small. Combining this observation with 
the application of low-overhead circuits for fault-tolerance may apply
to {\it every}  low-rate 
noise model for which the noise is invariant under permutations 
of the qubits. (Since constants in the depth translate to 
exponents in the overhead, such a result will not be practically useful.)



Complexity-theoretic reductions 
appear difficult and so is Scott Aaronson's nice challenge 
of a ``Sure/Shor separator'' 
\cite {Aa1}. 
A more realistic 
goal would be 
to prove that models of noise satisfying our conjectures do not allow 
for quantum linear error-correction, e.g., by deriving relations 
(\ref {e:p1}) and (\ref{e:c}) for any form of ``protected qubits'' obtained by
linear quantum error-correction.


\subsection {Potential counterexamples from physics}
\label {s:phy}

\subsubsection* {Topological quantum computers and anyons
}

In the area of ``topological quantum computers'' \cite {K2,FKLW,Col} 
there is 
a beautiful and powerful ``bilingual dictionary'' between certain forms of 
combinatorial methods for fault tolerance 
and remarkable objects from physics.  
It is 
suggested that fault tolerance can 
be realized 
by 
{\it non-Abelian anyons}, which can 
be thought of as analogous to the 
physical realization of logical
bits in a digital computer that are very robust to noise. The two 
``languages'' of this dictionary are a combinatorial 
descriptions of a quantum error-correction
with $n$ qubits, and their physical realization as 
certain quasi-particles called   
``anyons.'' 
Only a small number of types of non-Abelian 
anyons are required to realize the full power of quantum computers.

Analyzing the stability of non-Abelian (and Abelian) 
anyons based on the assumption that 
the noise is ``local'' (statistically independent, as discussed above) 
reveals a remarkable phenomenon referred to as ``mass gap.'' Below a 
certain temperature the anyon is 
going to be extremely stable. (Low temperature translates 
to low error rate.) The mathematical model 
predicts that as $n$ grows the region of stability (in terms of  
temperature) will not become smaller and
the ``gap'' will be maintained. 
Moreover, in this stable area the robustness 
to noise will be exponential in $n$ and 
thus, on the physics side, we will obtain
very robust qubits.\footnote {In the translation between 
a discrete combinatorial model with $n$ qubits
(or $n$ elements)  and a concrete physical object, 
it is unclear what the interpretation is on the physics side for
the value of $n$, and the relevance of the 
behavior as $n$ tends to infinity should not be taken for granted. 
Inner dependencies of the physical object appear to be relevant to the 
best value of $n$ for the combinatorial model describing it.
}

The existence of 
very robust (``protected'') qubits 
based on quantum error-correction via a highly entangled state, 
whether implemented 
by ``software,'' say  ion traps, or by ``hardware,''
say 
anyons, 
runs counter to our conjectures.
We can expect that when we study the effect of noise 
for the combinatorial 
model of anyons with a highly entangled state, using a perturbation method 
that reflects Postulates [P1] and [P2], the exponential robustness with $n$,
or even the ``mass gap,'' will disappear.

\subsubsection *{Other possible counterexamples from physics}

{\bf Superconductivity.} Several people have suggested 
that our postulates, and especially Conjecture [C], are already in 
conflict with 
phenomena from physics, 
like 
superconductivity 
and Einstein--Bose condensation. 
Superconductivity and related phenomena
are indeed physical systems with strong forms of (pairwise) entanglement 
that appear to be related to what is required for quantum fault-tolerance.

I tend to think that the form of entanglement for superconductivity  
is insufficient to refute 
Conjecture [C] since the entanglement in this case 
is ``generated'' (to a large extent) by 
dependencies of pairs of elements. 
Translating  and testing 
Postulate [P1'] for the setting of superconductivity would be of interest.

{\bf Cluster states.} 
Cluster states, defined by 
Briegel and Raussendorf (see \cite {RBB}) are roughly  
quantum analogs of (low-temperature) probability distributions 
described by Ising and Potts models. There is some simulation-based evidence 
\cite {GRAC} that certain materials from solid-state physics exhibit
a similar form of entanglement. Those materials can thus be potential 
candidates for checking empirically our conjectures on decoherence. On 
the other hand, the possibility of having universal quantum 
computation \cite {RBB}, and even fault-tolerant quantum 
computation \cite {ND}, based on cluster states and 
single-qubits measurements 
may challenge the relevance of our 
conjectures on the FTQC hypothesis.\footnote {What needs to 
be examined in this respect 
is the translation of 
non-Markovian  noise models (like those satisfying our conjectures) 
on the cluster-state computer, 
back to the quantum circuit it simulates; see \cite {ND}.}
For cluster states $\widetilde{ENT}$ appears to be linear in the number 
of qubits.

{\bf ${\bf 2n}$ bosons.} Noisy quantum computers 
that respect our postulates are incapable 
of simulating hypothetical objects like 
non-Abelian anyons. But are they capable
of simulating familiar, much simpler, objects from physics?
A simple example to test the conjectures  
is to consider them for a state $X$ of $2n$ bosons
($n$ large) each having a ground state $|0>$ and an excited state $|1>$,
so that $|0>$ has occupation number
(precisely) $n$ and $|1>$ has occupation number $n$. (A similar state $Y$
where the occupation number has a binomial distribution 
can be simulated by a tensor product state.) 


\subsubsection *{Other possible relations to  physics}

An obvious connection to physics is that 
a failure of computationally superior 
quantum computing 
would suggest 
that computations of quantum physics that are relevant to physical reality 
can  efficiently be simulated on classical computers,
and thus would question the relevance to reality
of computations from quantum physics
that appear to be computationally hard. We mention two other potential 
connections.

{\bf Perturbation methods.}
Another connection to physics may come from the perturbation methods 
used to analyze non-Abelian anyons. These methods 
are  related to standard perturbative
methods used in various other areas of 
physics and mainly in quantum field theory. 
Modifications of the 
perturbation method itself, which 
may amount to amending unjustified hidden 
probabilistic assumptions
and may lead to a drastically different behavior for the extreme
situation of (hypothetical) highly entangled systems 
like quantum error-correcting code and quantum computers, may be of interest
also in more mundane situations from physics, where these perturbation 
methods are (rather successfully) used.

{\bf Thermodynamics.}
Connections between fault tolerance and thermodynamics were considered,
e.g., in \cite {ALZ,AH}, 
and were 
intensely debated. (The results and methods of 
\cite {ABIN} also have a clear thermodynamic flavor.) 

For example, in a very recent paper, Alicki and Horodecki 
\cite {AH}  propose the following line of thought: 
1. Thermodynamics is relevant because
very robust storage of quantum information 
requires large systems. 2. Meta-stable states for 
finite systems are necessarily manifested by equilibrium states of 
infinite systems. 
3. Equilibrium states of infinite systems must have 
the form ``probability measures over a set of states'' and, in particular, 
cannot support even a single (``viable'') qubit. 
Of the above points, the second 
is perhaps the most controversial, and 
in view of some potential counterexamples 
may be related to 
noise models and perturbative methods that are different from 
the standard ones.   

The information-theoretic form of the mathematical formulation
of our postulates (which, to a large extent, were required 
in order to respond to various 
points discussed in this section) suggests possible connections 
with thermodynamics. Of particular interest are connections with entropy-type
measures of ``high order'' statistical dependence.  


\subsection{Conclusion} 
My belief is that the interesting question of the
physically realistic ``Church--Turing 
thesis'' 
(put forward mainly by
Deutsch) and, in particular, the feasibility of computationally 
superior quantum computers will have a convincing solution, and that, 
whatever this solution is, 
the asymptotic approach --- namely, the relevance of 
the asymptotic behavior of complexity to real-life 
computation --- will prevail. 


The question ``How can (computationally superior) 
quantum computers
fail''\footnote {While the possibility of 
computationally superior quantum computers 
certainly captures the imagination, it is worth noting 
that implementing even simple computations
on quantum systems 
can be important for 
applications, such as enhancing the performance of 
medical NMR \cite {BMRNL,Sch}.}
 is  as important a part of the quantum information and quantum
computers endeavor, as  the question ``How can (computationally superior) 
quantum computers succeed.'' As a matter of fact,
the two questions are the same.




\end {document}